\documentclass{llncs}
\pdfoutput=1
\usepackage{graphicx}
\usepackage{subfigure}
\usepackage{cite}
\usepackage{color}
\definecolor{darkred}{rgb}{1,0,0}
\definecolor{darkgreen}{rgb}{0,0.5,0}

\usepackage{soul} % To cross out: \textst

\begin{document}
\frontmatter          % for the preliminaries
\pagestyle{headings}  % switches on printing of running heads
\addtocmark{Semantic annotation of bioinformatics workflows} % additional mark in the TOC
\title{Automatic annotation of bioinformatics workflows with biomedical ontologies}
\titlerunning{Semantic annotation of bioinformatics workflows}  % abbreviated title (for running head)
\author{Beatriz~Garc\'{i}a-Jim\'{e}nez \and Mark D. Wilkinson}
\authorrunning{Beatriz~Garc\'{i}a-Jim\'{e}nez et al.} % abbreviated author list (for running head)
\tocauthor{Beatriz~Garc\'{i}a-Jim\'{e}nez and Mark D. Wilkinson}
\institute{Biological Informatics Group\\
Center for Plant Biotechnology and Genomics (CBGP), UPM - INIA\\
28223 Pozuelo de Alarc\'{o}n (Madrid), Spain\\
\email{beatriz.garcia@upm.es, markw@illuminae.com}\\
\texttt{http://www.wilkinsonlab.info}
}

\maketitle              % typeset the title of the contribution

\footnote{\textit{\textbf{Thanks to Springer. The final publication is available at link.springer.com}}}

\begin{abstract}
Legacy scientific workflows, and the services within them, often present scarce and unstructured (i.e. textual) descriptions. This makes it difficult to find, share and reuse them, thus dramatically reducing their value to the community.  This paper presents an approach to annotating workflows and their subcomponents with ontology terms, in an attempt to describe these artifacts in a structured way.  Despite a dearth of even textual descriptions, we automatically annotated 530 myExperiment bioinformatics-related workflows, including more than 2600 workflow-associated services, with relevant ontological terms. Quantitative evaluation of the Information Content of these terms suggests that, in cases where annotation was possible at all, the annotation quality was comparable to manually curated bioinformatics resources.
 
\keywords{scientific workflows, web services, bioinformatics, semantic annotation, text mining, ontologies, tags, term extraction}
\end{abstract}

\section{Introduction}
As the demand grows for more transparent and reproducible scientific research \cite{Micheel2012}, it becomes increasingly urgent to adopt more formal strategies for recording scientific methodology.  The most common approach to such explicit process modelling takes the form of a scientific workflow.  These digital artifacts formally describe the series of steps by which a scientific experiment was/will be conducted.  Workflows may exist at a variety of levels of abstraction, ranging from general process overviews, to specific tools, the data-flow connections between them, and their associated execution parameters.

% B:  we could probably remove this entire paragraph, so long as we move the relevant references to other places in the text... 
Formal workflows are generally authored and/or executed using purpose-built software, and a variety of design and enactment environments are used by e-scientists, such as Taverna \cite{Oinn2004}, Kepler\cite{altintas2004kepler}, Wings \cite{wings2011}, and Vistrails \cite{callahan2006vistrails}.
These environments serve three main purposes:  first, they attempt to generalize the interfaces between different workflow components (e.g. Web Services versus local command-line tools or scripts) so that they can be connected together without concern for the precise mechanism by which data will be passed between the components; second, they often offer a means to facilitate component discovery at design-time, either by menu-driven component selection \cite{Oinn2004}, by contextually-aware suggestion \cite{Withers2010}, or by semi or fully automated construction \cite{Vandervalk2009b} \cite{wings2011} to simplify the design process and/or reduce errors; finally, at enactment time, they mediate the data flow between components and, generally, capture additional information about the provenance of the workflow execution.

Apart from the key goal of enhancing the explicitness and transparency of scientific methodology, one of the most touted benefits of formal workflows is that they can, in principle, be shared, reused, and repurposed.  With this goal, and in parallel with the increasing use of formal workflows in e-science, projects have emerged that aim to capture and publish these workflows for the purpose of rediscovery and reuse.  The primary such repository in the Life Sciences is myExperiment \cite{bib:myExperiment-bioinfo}, which archives workflows from most design and enactment environments.

To facilitate discovery, workflows submitted to myExperiment can be annotated with both a block of descriptive text, as well as free-text keywords or \textquotedblleft tags"; however there is little to no control over the quality or quantity of these annotations.  Task-appropriate workflow discovery, then, relies largely on the matching of keywords from within these freeform, sometimes very limited textual sources.  
Similarly, detailed comprehension of the functionality and suitability of a discovered workflow also depends largely on human interpretation of these textual annotations.  For example, if the workflow requires edits for repurposing, deep examination of the individual workflow subcomponents is required in order to identify which portion of the workflow requires revision.  Unfortunately, workflows are seldom, if ever, annotated at this level of granularity.  As such, it becomes necessary to resolve the individual subcomponents to their own sources of documentation.   
Such documentation might be available, for example, within the WSDL document for a Web Service, or the record of that service in BioCatalogue, both of which are, again, either freeform tags or narrative text.  While such traversals are plausible, there is currently no infrastructure that can reliably mechanize the traversal from a workflow subcomponent to its independent documentation.  Moreover, the documentation of these subcomponents is as unregulated and often as sparse as that of the workflow itself, thus making them of dubious utility.

One approach to improving the status quo would be to semantically annotate both workflows and their subcomponents with ontological terms. Semantic annotations have numerous benefits over keyword and free-text annotations, such as supporting query expansion, filtering, precision, and computational tractability for formal verification and validation of workflow structures.  
There are currently no widely accepted standards for representing or annotating workflows though a variety of \textit{de facto} standards are available; conversely, more widely accepted standards are available by which to capture semantic annotations for individual workflow components.  For example, the World Wide Web consortium has recommended the SAWSDL standard for capturing ontological and controlled vocabulary terms within the structure of a traditional WSDL document representing a Web Service.  This is intended to act as a bridge between traditional Web Services, and \textquotedblleft Semantic Web Services", where each field in the conventional Web Service interface definition can now be mapped into an ontological context, together with an (optional) machine-readable data structure mapping, such as XSLT.
Projects such as EMBRACE \cite{rice2006embrace} are taking on the task of annotating legacy Web Services into SAWSDL using ontological terms from bioinformatics ontologies, in particular, EDAM (EMBRACE Data And Methods ontology \cite{bib:EDAM}).  Other projects aim to take advantage of even richer semantics.  The SADI \cite{Wilkinson2011}, SHARE  \cite{Vandervalk2009b} and Wings  \cite{wings2011} projects all capture rich semantic annotations at the level of both the overall workflow, as well as the individual subcomponents, and hence, both support (to some degree) fully automated workflow assembly.  In all of these cases, however, the semantic annotations are generated manually, either at the time of service/workflow authoring, or as part of a legacy migration and curation process.  As such, it would be highly desirable to \textquotedblleft boot-strap" the semantic annotation of legacy workflows and workflow subcomponents through some form of automated semantic annotation.

Here we describe an approach to the semantic annotation of legacy workflows in the myExperiment repository, as well as their component services.  Workflows are first filtered to eliminate any steps that are exclusively syntactic transformations (\textquotedblleft shims" \cite{Radetzki2006}), with the resulting workflow skeleton containing only \textquotedblleft biologically meaningful" operations. 
These skeletons are then mined using information from a variety of sources, including the myExperiment and BioCatalogue \cite{Bhagat2010} repository entries, the BioMoby registry \cite{Wilkinson2008}, and WSDL source documents.  Mined descriptions are then processed to discover matches to nine relevant ontologies from the OBO Foundry.  The resulting workflow templates are then reassembled using the representation of the Open Provenance Model for Workflows (OPMW) \cite{bib:OPMW-2011,bib:OPMW-2012-report} from the Wings project, which includes well-defined facets for capture of rich semantic annotations.  Finally, we describe our degree of success in extracting such annotations, as well as attempting to quantitatively evaluate the quality of these annotations in terms of their Information Content \cite{bib:IC-Sanchez2011}.

\section{Material and Methods}
\label{sec:Methods}
Section \ref{sec:Methods} describes the steps we undertook in our efforts to automatically annotate myExperiment workflows using terms from bioinformatics-relevant ontologies. 

Overall, the system consumes Taverna workflows in Taverna 1 (\textit{scufl}) or Taverna 2 (\textit{t2flow}) formats \cite{bib:taverna-2004}, and outputs a set of ontology annotations linked to each available and \textquoteleft biologically meaningful' service within each input workflow. As a secondary output, the partially-abstracted workflow (i.e. without data-type transformation nodes) is provided in \textit{scufl} or \textit{t2flow} format.

The following subsections describe the four primary phases of our analytical approach: 1) Filtering for bioinformatics-relevant workflows, 2) Cleaning \textquotedblleft shim" services, 3) Retrieving service descriptions and 4) Entity extraction from descriptions to create the output semantic annotations.

% Without enough pages to include these figure.
% Fig. \ref{fig:schemaWfSystem} illustrates each step described in this section.
% 
% \begin{figure*}%[!tpb]
% 	\centerline{
% 	%	\includegraphics[scale=0.75]{images/schema_wfsystem.pdf}
% 	}
% 	\caption{\textbf{Schema of our system that annotates Taverna workflows with bioinformatic ontology terms.} The figure shows the input data, output data, activities and connections of the different steps described in section \ref{sec:Methods}.
% 	}
% 	\label{fig:schemaWfSystem}
% \end{figure*}

\subsection{Step 1: Filtering for Bioinformatics-Relevant Workflows}
Our first requirement was to differentiate bioinformatics-oriented workflows from those relevant to other areas of investigation.  As a first attempt, we selected workflows with the \textit{\textquoteleft Bioinformatics'} tag; however, we observed that only 5\% of Taverna workflows are described with this tag, leading to a high false-negative rate. % 2014.02.26: 94 of 1826 Taverna (1 and 2) workflows.
We adjusted our criterion to search for specific bioinformatics-oriented topics, using relevant branches of EDAM \cite{bib:EDAM} as our source vocabulary.   Relevant EDAM terms were derived using the \textit{\textquoteleft edamdef'} command from the EMBOSS package v6.4.0-4 \cite{bib:EMBOSS}, % [Probably, it uses EDAM v1.2 or earlier (I think so)]
which allows us to search the definition of EDAM classes and returns terms matching the query term(s).  The query:

 \texttt{edamdef -namespace topic -subclasses -query bioinformatics}
 
\textit{\textquoteleft edamdef'} returned 190 terms related to \textquoteleft bioinformatics' from the \textquoteleft topic' sub-ontology of EDAM.  The description, title and tags of each myExperiment workflow were then searched using each of these terms, using the text mining Peregrine SKOS CLI software \cite{bib:Peregrine}.  This filtering process resulted in 1206 presumptively bioinformatics-related workflows.

From manual inspection of these 1206 workflows, it became apparent that there were still an unacceptable number of false negatives because of the lack of an EDAM term in the description, title, or tags.  Importantly, it was also apparent that some selected EDAM terms were too general, resulting in an unacceptable number of false positive workflows. For example, many false positives were discovered by matching the EDAM term \textquoteleft workflows' (1022 cases, mixed with true positives) or \textquoteleft ontologies' (29 cases).

After several iterations of trial and error together with manual verification of filtered and non-filtered workflows, we curated the list of filter terms, removing many of the most general EDAM classes which select workflows not specifically related to bioinformtics (e.g. ontologies, rna, structure, text mining, threading and workflows) and adding new specific terms to include bioinformatics workflows not retrieved with the EDAM classes (e.g. alignment, bioinformatics, BioMarker, bioMart, BioMoby, blast, chEBI, chemical, cheminformatics, EBI, ebi.ac.uk, ensembl, entrez, FASTA, GenBank, Gene expression, gene list, gene name, Gene Ontology, gene pattern, geneontology, genetic, genotyping, GO term, InterPro, Kegg, metagenomics, microarray, molecular, molecule, ncbi, openPHACTS, pathway, Pfam, phylogenetic, protein, PubMed, SNP, somatic, SwissProt, systems\_biology uniprot, UniprotId and wikipathways).  We then repeated the filtering process.  Again, a manual examination of a subset of the filtered workflows suggested that approximately 95\% of the erroneous filtering had been eliminated; the identified false positives had been eliminated, preserving the true positives, and 
many of the false negatives were now discovered. Remaining false negatives consisted of workflows with no description, no tags, no title and/or words not correctly space-separated \textemdash{}effectively, impossible to discover using our approach.  We believe, however, that this set of terms provides sufficient filtering precision to be used in an automated annotation pipeline leading to a dataset of sufficiently high-quality to be used in downstream data mining.

At the end of this filtering phase, from an input of 1839 workflows, 775 workflows were determined to be relevant to bioinformatics, although just 739 workflows are available to download from myExperiment (4.65\% not downloadable).  Among the 739 available workflows, 272 of them are in Taverna 1 format (scufl) and 467 in Taverna 2 format (t2flow).

\subsection{Step 2: Cleaning \textquotedblleft shim" Services}
Taverna workflows have been reported as containing many \textquotedblleft shim" services \cite{bib:TavernaMotifs-2012} \textemdash{}that is, workflow elements that execute data transformations (merging, formatting, or parsing), but not biologically meaningful analyses \cite{bib:taverna-2013}. These shims represent structural transformations, not biologically relevant transformations we are interested in. They do not contribute to our understanding of the science behind a workflow, and as such, we undertook to automatically identify and remove them from the workflow prior to the annotation phase of our analysis.

\begin{figure*}[hbt*]
	\centering
	\subfigure[Before]{
		\includegraphics[scale=0.25]{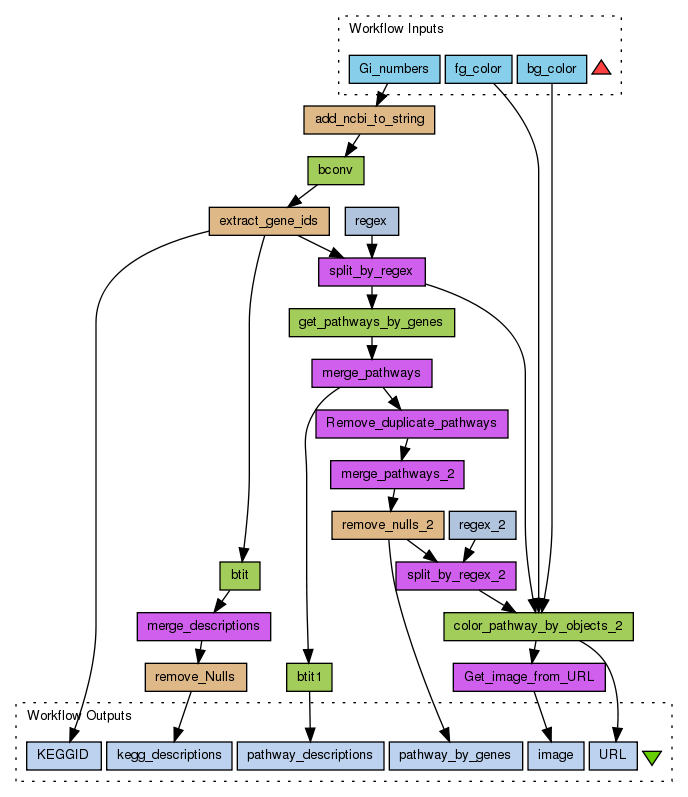}
		\label{fig:cleanedShimsExampleBefore}
	}
	\subfigure[After]{
		\includegraphics[scale=0.4]{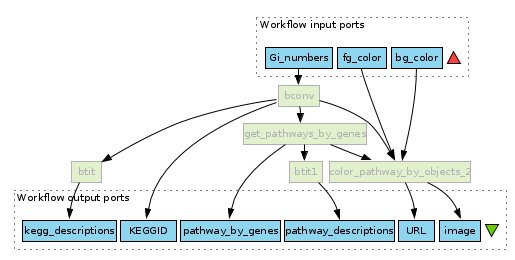}
		\label{fig:cleanedShimsExampleAfter}
	}
	\caption{\textbf{Example of workflow with cleaned shim services.} \subref{fig:cleanedShimsExampleBefore} The original workflow is myExperiment workflow \#1180, and \subref{fig:cleanedShimsExampleAfter} its associated service without shims.
	}
	\label{fig:cleanedShimsExample}
\end{figure*}

According to the Taverna User Manual\footnote{\textit{http://dev.mygrid.org.uk/wiki/display/taverna/Service+types}}, we considered as shim services the following categories: XML splitter, spreadsheet import, string/text constant, beanshell, local service and Xpath; and as non-shim services: WSDL, REST, bioMoby, bioMart, soaplab and Rshell.
When a shim service is removed, the steps before and after that shim are reconnected in our dataset, thus preserving the \textquotedblleft flow" of the workflow.  Note, however, that the resulting T2flow files cannot be accurately visualized in Taverna (though they appear to be XML schema-compliant); nevertheless, since visualization was not our objective, nor was the objective to create a \textquotedblleft runnable" workflow, this was not problematic for the remainder of our analysis.
% \begin{table}
% 	\caption{\textbf{Classification of categories of services between shim or domain services.} According to the Taverna User Manual (\textit{http://dev.mygrid.org.uk/wiki/display/taverna/Service+types}).}
% 	\label{tab:shimServices}
% 	\begin{center}
% 	  \includegraphics[scale=0.38]{images/tableShims.jpeg}	  
% 	\end{center}
% \end{table}
In some cases, the pruned workflow has services without inputs and/or outputs, and in other cases, pruned workflow is left with only inputs and outputs, if all its processors are shims.  
Figure \ref{fig:cleanedShimsExample} shows an example of a Taverna workflow before and after cleaning shims.

At the end of this second data preparation phase, we retain the same number of workflows overall, however each workflow now has fewer component processors.  77 workflows contained only shim services, and were therefore \textquotedblleft empty" after this data preparation phase.

\subsection{Step 3: Retrieving Service Descriptions}
Here, we query myExperiment and a variety of service metadata repositories to obtain a textual description of each remaining service in each workflow.  We focus our efforts on annotations present in WSDL, BioMoby, SoapLab, REST and nested workflows services, since they are the most frequent services in our workflows (see section \ref{sec:results-WFcomposition}) and have obvious metadata sources within which to search for annotations.

For each service, we attempt to construct a textual description that is composed of (if available): service name + service description + operation name + operation description. These four different elements are discovered from a variety of sources and sites, using several keys (e.g. endpoint URI or service name), with continuous checking for errors, and with multiple, possibly redundant attempts, attempting to locate the richest, most descriptive source possible until a description is found or all possibilities have been expended.  
The sources included any or all of:  the myExperiment workflow entry; Scufl and T2flow files from myExperiment \cite{bib:myExperiment-bioinfo}; the WSDL source document for each service; the BioMoby registry entry \cite{Wilkinson2008}\footnote{\textit{http://moby.ucalgary.ca/cgi-bin/getServiceDescription}}; and the BioCatalogue service repository \cite{Bhagat2010} through its API to specific endpoint searches and general searches. For Scufl files, service descriptions were sometimes available for services within these files. T2Flow files do not have a descriptive field, other than for nested workflows, and as such it was always necessary to attempt retrieval of the WSDL source document, MOBY registry entry, or retrieve the relevant record from the BioCatalogue repository.

Disappointingly, very frequently, a dearth of annotations at the service level meant that the final textual description of a service was limited to just the service name (912 of 3560 bioinformatics services - 25.62\%) with 246 services having no annotation whatsoever.
%and among the 3314 services with \textit{some} description, 45.16\% of them fell into this category.

\subsection{Step 4: Entity Extraction from Descriptions to Create Semantic Annotations}
\label{sec:Methods-step4}
The final step in our annotation pipeline is to execute text analytics on the description from step 3, in order to ontologically annotate the services and, subsequently, the workflow of which they are component. As such, the input to this step in the pipeline is a descriptive paragraph, and the output is a list of relevant ontology classes associated with that service description.

The ontologies that acted as the vocabulary source for the text analysis were:  BioAssay Ontology (BAO), %\cite{bib:BAO}, 
Bioinformatics Web Service Ontology (OBIWS), % \cite{bib:OBIWS},
Biomedical Resource Ontology (BRO), % \cite{bib:BRO},
EDAM Ontology of Bioinformatics Operations and Data Formats (EDAM), % \cite{bib:EDAM},
Experimental Factor Ontology (EFO),
Information Artifact Ontology (IAO), % \cite{bib:IAO}, 
Mass Spectrometry Ontology (MS), % \cite{bib:MS},
Medical Subject Headings (MESH), % \cite{bib:MESH}, 
National Cancer Institute Thesaurus (NCIT),
Neuroscience Information Framework Standard (NIFSTD),
Ontology for Biomedical Investigations (OBI), % \cite{bib:OBI},
Semanticscience Integrated Ontology (SIO) and
Software Ontology (SWO). % \cite{bib:SWO}

These ontologies cover a variety of categories of concepts relevant to bioinformatic workflows, such as operations, topics, algorithms, etc.  All of them must belong to the OBO Foundry \cite{bib:OBOfoundry} and BioPortal \cite{bib:BioPortal} library of ontologies, and thus can be used for annotation using the \textit{Open Biomedical Annotator} \cite{bib:OpenBiomedicalAnnotator}.  The \textit{Open Biomedical Annotator} is an application available from BioPortal \cite{bib:BioPortal}, an open repository of commonly used biomedical ontologies and related tools.  The \textit{Open Biomedical Annotator} web service matched words in our descriptive paragraph to classes in selected ontologies by doing an exact string comparison (a \textquoteleft direct' match) between our description and ontology class names, synonyms and identifiers.  %http://bioportal.bioontology.org/help?pop=true#Annotator_Tab
 
In some cases, we noticed duplicated annotations (with the same or different URI), due to overlapping or explicit relations among different ontologies (e.g. SWO imports EDAM). In section \ref{sec:results-AnnotAnalysis} we present results referring to the total numbers of annotations before and after removing duplicated terms, where duplication is defined as sharing the same URI, but appearing in different ontologies. To remove the redundancy, we consider SWO, OBIWS, OBI, EFO and NIFSTD have preference to their imported ontologies. Terms with the same name/label, but differing URIs, are \textit{not} considered to be identical, since that would require a deep, manual interpretation of the semantics of that term within each of the ontologies.

530 workflows from the 739 available bioinformatics workflows were successfully annotated with one or more ontological classes.

\section{Results}
First, this section presents a study of the workflows in terms of various categories of sub-component composition. Subsequently, we expose an analysis of the quantity and quality of the derived semantic annotations, based on a calculation of the Information Content represented by the set of semantic terms discovered. Finally, a sample of the final workflow annotations represented according to the OPMW model is provided.

\subsection{Understanding Workflow Composition}
\label{sec:results-WFcomposition}
\begin{figure*}%[!tpb]
	\centerline{
		\includegraphics[scale=0.50]{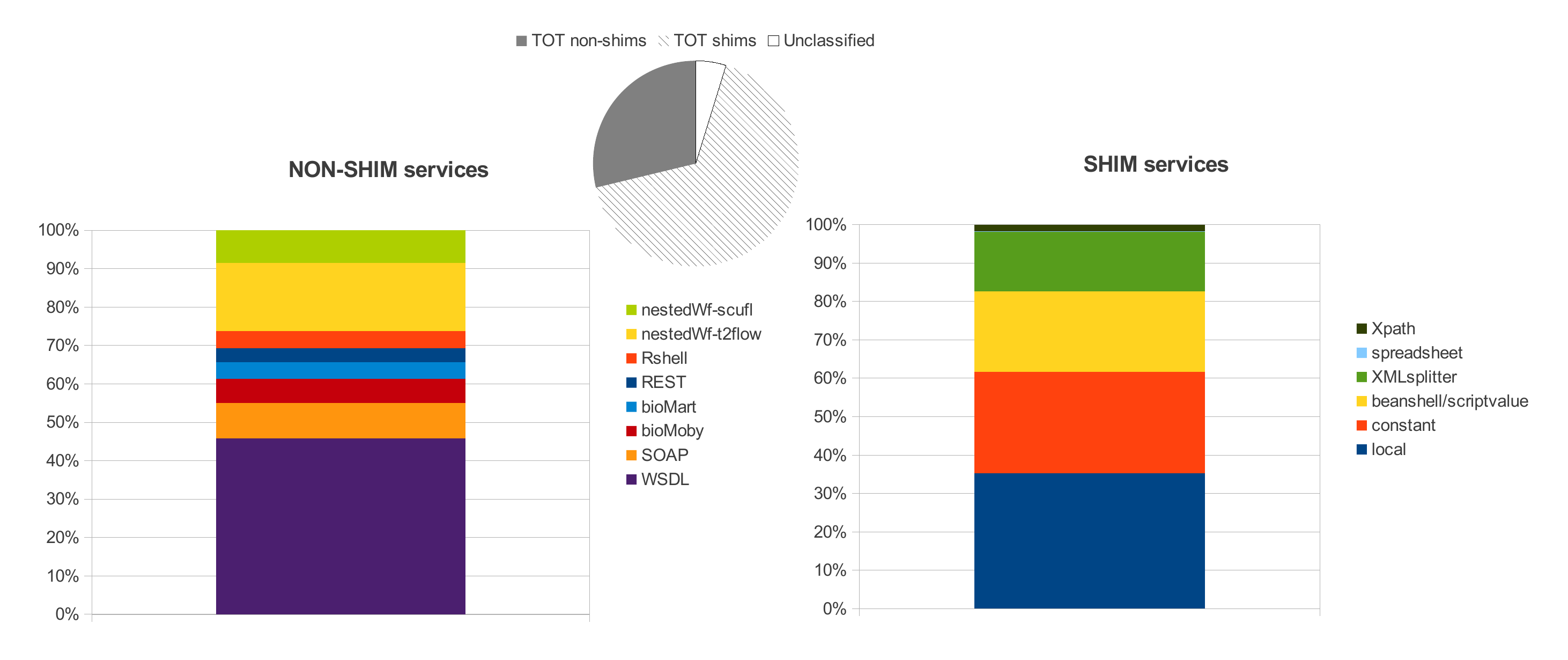}
	}
	\caption{\textbf{Average distribution of shim and non-shims services.} In the center, the global average distribution between shim and non-shims services in all bioinformatic workflows. On both sides, the average distribution of shim (right) and non-shim (left) categories of services.
	}
	\label{fig:shim-NonShims}
\end{figure*}

Figure \ref{fig:shim-NonShims} illustrates the distribution of the different categories of services in all 739 bioinformatics workflows, split into shim and non-shim components. The most notable observation in Fig. \ref{fig:shim-NonShims} is that the number of shim elements far exceeds the number of biologically meaningful elements; more than 65\% of all workflow components are shims (see Fig. \ref{fig:shim-NonShims}(center)). This reinforces the importance of step 2 of our annotation system, but also highlights the complexity and penetrance of data transformation problems in bioinformatics, at least in part due to the proliferation of data formats, as has been argued for at least a decade \cite{Lord2004}.

Among the non-shim services (see Fig. \ref{fig:shim-NonShims}(left)), the most frequent are WSDL services, where, together with SOAP and BioMoby services, these make-up 60\% of all biologically-meaningful services.  The most common categories of shim services (see Fig. \ref{fig:shim-NonShims}(right)) are local, constant and beanshell/scriptvalue services, covering more than an 80\% of all shims combined.

On average, each bioinformatic-related workflow has 16.74 components, where 11.13 are shims and 4.81 are non-shim services. The remaining 0.80 are other unclassified services.  We observe that the ratio of shim/non-shim services is higher in T2flow format (12.50/4.81) than in Scufl format (8.77/4.83), where more than 2.5 shims are included per each domain service. % 2.59 vs 1.81 (scufl)

\subsection{Annotation Analysis}
\label{sec:results-AnnotAnalysis}
In the final step of our analysis, we attempt to quantitatively evaluate the degree to which this annotation methodology yielded useful results. 

One immediate consideration is that not all the services and workflows were amenable to automated annotation at all (i.e. had little or no information to mine for annotations).  We consider this to be a failure of the scientific community, rather than a failure of our analysis, and as such, we take this into-account with respect to our final evaluation of the pipeline's success.  

%First: Proportion of annotated workflows and services (according to totals in previous section).
In total, we obtain 70636 ontological annotations spanning 2922 different ontological classes, of which 64324 are non-redundant (i.e. the identical URI appearing in multiple ontologies). This means we achieve at least one annotation for 2605 of 3560 non-shim services (73.17\%). In terms of workflows, we collect the annotations of their instantiated services, without taking the order of the services into account. This allowed us to annotate 530 out of 739 workflows (71.72\%) with at least one ontology class.
% Include if I get to compute these numbers!!!!!!!:
% This results give an average of EEEE annotations/service, with a high standard deviation of FFFF. It is due to the high number of not annotated services and to the big difference in the number of annotations we can get in distinct services, according to the length and the quality of their corresponding descriptions.

Clearly, not all semantic annotations are equally informative.  For example, the ontology term \textquotedblleft analysis" is less informative than the ontology term \textquotedblleft fastq file parser".  Thus, we attempted to measure the informativeness and the relative quality of our automatically-generated annotations.  To achieve this, we apply a semantic metric based on Information Theory \textemdash{}the Information Content (IC) \cite{bib:IC-Sanchez2011}\textemdash{} and we computed the IC value of each automatically-selected ontology term, each service, and each workflow.

We choose an intrinsic IC metric for a variety of reasons, such as the topology of the taxonomy, the lack of a \textquotedblleft gold standard" annotated knowledgebase, and in order to avoid biases and the dependence on external annotations.  Among the three available alternatives for intrinsic IC \cite{bib:IC-Sanchez2011}, we chose the Zhou et al. metric \cite{bib:Zhou-2008} which takes into account the number of descendants in a similar manner to that proposed by Seco et al. \cite{bib:Seco-2004}, but where the former approach also includes the depth of a term in the taxonomy. Although Sanchez et al. \cite{bib:IC-Sanchez2011} improves this statistic by including the number of subsumers, we prefer Zhou et at. since they define a normalized metric, which therefore allows us to compare IC values of terms from distinct ontologies.

\begin{figure*}[!tpb]
	\centerline{
		\includegraphics[scale=0.6]{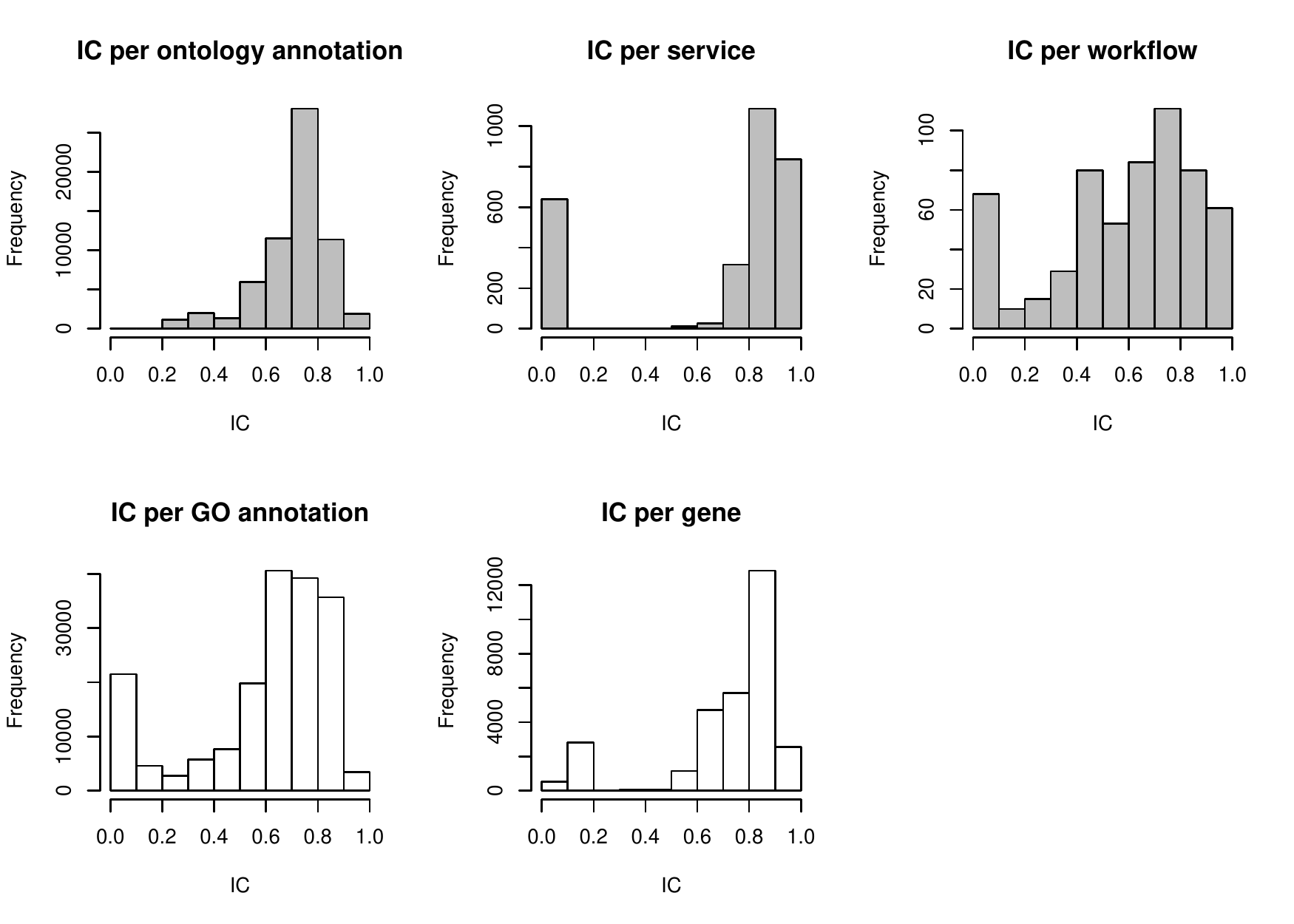}
	}
	\caption{\textbf{Histograms of Information Content (IC) values associated to ontological annotations.} Top row (grey columns) shows IC distribution of automated annotations generated by our system. Bottom row (white columns) shows IC distribution of manual GO annotations of \textit{Arabidopsis Thaliana}. Frequency axes are not directly comparable.
	}
	\label{fig:IChistogram}
\end{figure*}

Using the Zhou et al. metric, we compute the IC values of each ontology class with the Semantic Measures Library Toolkit \cite{bib:SMLtoolkit}. Thereafter, we describe the \textquoteleft informativity' of the set of semantic annotations associated with a service as the IC for that service, computed as the best (the maximum) IC value of every ontology annotation associated to that service. This measure is independent of the redundancy present within the annotation terms, since it takes only the maximum for any given term. Finally, we defined the degree of informativity of the annotations of an entire workflow, as the IC per workflow, computed as the average of all the IC values of the services within the workflow.  The top row in Fig. \ref{fig:IChistogram} shows the resulting IC value distribution of those automatic annotation. Note that only services that had some analysable description are included in these IC values computations, since the others could not be scored.
Other reasons for not being scorable include annotation with obsolete terms, or terms not included in the main ontology hierarchy since they are external properties associated with a core ontology class, and therefore not part of the IC statistic, such as terms from MESH qualifiers and MESH supplementary concept records.

Having selected an intrinsic metric for IC, we were consequently unable to objectively define what a \textquotedblleft good" IC value would be.  Moreover, we do not have access to a \textquotedblleft gold standard" set of annotated bioinformatics services with which to compare our automated annotations.  As such, we opted to execute our IC analysis on a set of ontological annotations done largely manually, and considered to be of high-quality by the community. In particular, we take the set of genes of \textit{Arabidopsis thaliana} manually annotated with Gene Ontology (GO) terms from the TAIR FTP site \cite{bib:TAIR}. We compute the IC values for each gene, as the IC of the set of GO terms associated with that gene, using the same procedure as with our automatic annotations; thus, we infer a hypothetical correspondence between an annotated locus, and an annotated service.  The results of these IC values corresponding to manual annotations is shown in bottom row of Fig. \ref{fig:IChistogram}.

When both rows are compared in Fig. \ref{fig:IChistogram}, we observe a similar distribution of IC values resulting from our automated annotations (top row) compared to the (largely) manual annotations (bottom row), with the notable exception of the high number of services with annotation IC of 0 (due to a lack of annotations). On average, the IC value per automated annotation is 0.7139 versus 0.5986 for the manual GO annotations (first column) and the average IC value per annotated service is 0.8707 if all IC=0 services are excluded 0.6801 if they are included) versus 0.7172 of manual annotation per gene. We note that genes of unknown function (5229 Arabidopsis genes) are excluded from the GO annotation file, in a manner similar to our filtering-out of services with no annotation, increasing the validity of this comparison. %These are either genes of unknown function, microRNAs (also of unknown function) or pseudogenes}.
Therefore, we could conclude that the informativeness, in terms of IC values, of our automatic annotations are as good or better than what we might expect from manual annotations.

To compare IC values split by ontology, we compute IC per ontology as the average of all the annotations within each ontology. We conclude SIO provides the best annotations (0.8172), followed by NCIT (0.7529) and SWO (0.7519); additionally, MESH has the highest minimum IC value (0.3459). In terms of quantity of annotations, NCIT is the best with 31853 annotations (1544 different terms).
The ten most frequent annotations were (in descending order): \textit{job resource} (NIFSTD), \textit{computer job} (NCIT), \textit{occupation} (NCIT), \textit{gene/s} (MESH and NCIT), \textit{protein} (EFO and NCIT) and \textit{database} (NIFSTD, MESH and EDAM).

\subsection{Workflow Annotations in OPMW Model}
To make our annotations available in a structured and reusable way, we chose to represent them in RDF as instances of the \textit{Workflow Template Process} class\footnote{\textit{http://www.opmw.org/ontology/WorkflowTemplateProcess}} from the Open Provenance Model for Workflows (OPMW) ontology \cite{bib:OPMW-2011,bib:OPMW-2012-report}.  While it may appear to be more desirable to publish these as SAWSDL documents, as recommended by EMBRACE, we elected not to do so, since (a) not all of our annotated services are originally published as WSDL and, moreover (b) there is no obvious way to re-publish the derived SAWSDL files in a way that would be discoverable/usable by the community (i.e. they cannot be re-associated with their respective services or workflows in either the BioCatalogue or myExperiment repositories).

Figure \ref{fig:OPMWtemplate} shows the OPMW structure that describes the semantic annotations of one service.  Our RDF output files\footnote{Available at \textit{http://wilkinsonlab.info/myExperiment\_Annotations/OPMW/*}}, includes an instance of this model for each non-shim available service of each annotated bioinformatics workflow.  Using this RDF file, our annotations are available and could be easily integrated in other systems requiring structured annotations of bioinformatic services.
\begin{figure}[hbt*]
	\begin{center}
	{\scriptsize
	 \texttt{
	\begin{tabbing}
\hspace*{0.25in} \= \hspace*{1.65in} \= \hspace*{0in} \kill
<SERVICE URL>\\
       \> a opmw:ProcessTemplate, <ontology class 1 URL>, <ontology class ...>, <ontology class N>;\\
       \> opmw:template <WORKFLOW URL>; \>\# link to workflow which this service is part of.\\
       \> opmw:uses <DATA URL>.         \>\# link to previous service in the workflow.\\
	\end{tabbing}
	} % texttt		
	} % scriptsize	
	\end{center}
  \caption{\textbf{Abstract structure to define our semantic annotations in one service}. }
  \label{fig:OPMWtemplate}
\end{figure}

In addition to this primary output, we also provide the derived set of partially-abstracted workflows (i.e. after removing non-biologically-meaningful steps) in Taverna formats\footnote{Available at \textit{http://wilkinsonlab.info/myExperiment\_Annotations/abstract\_workflows/*}} with the caveat that these are for informational purposes only, and cannot be accurately visualized, nor run, in Taverna.

\section{Discussion and Conclusions}

The limitations of this approach to bootstrapping annotations are obvious (both \textit{a priori} and  as borne-out in the results).  Namely, there was a well anticipated difficulty in finding descriptive annotations which could be mined for semantic meaning.  Beyond that, however, the heterogeneity of the content and representation of the workflows also made it difficult to discover, mine, and even select appropriate ontologies for the annotation effort.  We will now go into some details about how this affected the accuracy and/or comprehensiveness of each step in our annotation pipeline.

With respect to the first step \textemdash{}filtering for bioinformatics-related workflows\textemdash{} when a workflow is submitted to myExperiment, fields that could justifiably be considered \textquotedblleft core metadata", such as description, and title, are not mandatory.  The same can be said of service submission to the various Web Service registries.  Disappointingly, workflow and service submitters therefore can, and too often do, opt to disregard these most basic metadata elements.  Given that we depended on this metadata for our first-pass filter of relevance, we can be certain that our outcomes were adversely affected by these \textquotedblleft nuisance-behaviours".  In the absence of these basic information elements, it would be extremely difficult even for a human to determine the function and/or relevance of a workflow for a particular task, and it was clearly an insurmountable problem for an automated annotation pipeline.
Finally, one additional limitation to our success in the first stage of our pipeline was that not all selected workflows were available to download. MyExperiment returned \textquoteleft not found' or \textquoteleft not authorized' errors in these cases, and thus these workflows had to be removed from our analysis.

Lack of descriptive annotations became acutely problematic in step 3 of our pipeline, when we attempted to construct a description of individual services within the workflow.  It was frequently the case (25.62\% of \textquoteleft biologically relevant' services) that the files describing a service would have no description at all beyond the service name.  We would suggest that such lax, nonchalant behaviours on the part of submitting scientists entirely defeats the purpose of submitting to a public repository.  Therefore we believe that it would not be unreasonable for the various workflow and service registries to be more demanding of authors with regard to these fundamental annotations.

This same problem manifested itself in the semantic annotation retrieval from step 4, where the quantity and quality of the derived annotations depend, obviously, on the amount of descriptive text available; the longer the description, the more informative the annotations in most cases. For example, services with short descriptions (such as \textquotedblleft prophet: Scan one or more sequences with a Gribskov or Henikoff profile\textquotedblright, with 63 characters), result in few and very general ontology annotations (\textit{Sequence analysis} (EDAM) and \textit{scan} (MS), with a IC per service of 0.5785). While services with longer descriptions result often in many and specific ontology annotations (such as \textquotedblleft Eigen\_analysis\textquotedblright, with 3946 characters, with 263 annotations, with an IC per service of 0.9588).
% Longer service: wf3686

A distinct source of error arose from service deprecation.  Several services referred-to in workflows had been deprecated, and the sources of documentation (if they ever existed) were absent.  Although through manual exploration we determined that some of these deprecated services have been replaced by new services, and these new services had annotations in Biocatalogue, it was very difficult to automatically discover when such deprecation/replacement had taken place based on the reference to that service in the original workflow, and it was not clear how to automate this complex traversal.  It seems, therefore, that some clearly-defined method for tracking versioning is required for workflow sub-components, and that this tracking mechanism should have features that allow it to be automated.

Finally, existing annotations were missed due to external repositories' errors and bad (malformed or inappropriate) responses to search queries.  These cases, though representing only a fraction of the entries, were the result of either unstable interfaces and/or errors (or non-documented limitations) in the various APIs.

Finally, related to the quality of the automated annotations, we emphasise that IC value can only give a measure of the informativeness of the annotations; it cannot report on their appropriateness vis-\`{a}-vis the real function of the service or workflow.  Moreover, IC value is determined largely by the topology of the ontology, and because ontologies vary in their granularity from branch-to-branch, high IC values do not necessarily imply \textquotedblleft rich" annotations.  For example, some annotations with high IC values, such as \textit{patient, synonym, human} and \textit{length}, with IC scores greater than 0.95, do not seem, subjectively, particularly informative in the context of the aims of our study.
However, IC is an objective and useful measure when an ontology is applied to annotate a knowledge base \cite{bib:thesisBenGood} and we could not identify an objective alternative.

Altogether, we feel there is significant room for improvement in the straightforward capture of core, non-semantic metadata at the point of resource submission.  We demonstrate here how this, in turn, could allow automated semantic annotations of surprisingly high quality to be extracted via text mining approaches.  Such semantic annotations could then be used to improve the submission process itself, by for example, detecting when certain important types of metadata are missing, and/or prompting for likely annotations based on existing patterns detected by Machine Learning techniques.

\section*{Acknowledgments}
MDW is funded by the Isaac Peral/Marie Curie Cofund programme (FP7) of the European Union and UPM; BGJ is funded by the Isaac Peral Programme of UPM.

%\bibliographystyle{plain|unsrt|alpha|abbrv}
%\bibliographystyle{plain}
\bibliographystyle{splncs.bst}
\bibliography{references_isola2014.bib}

\clearpage
\end{document}